\documentclass[final, runningheads]{llncs}
\usepackage[T1]{fontenc}
\usepackage{graphicx}
\usepackage{amsmath}
\newcommand{\itadata}{\footnotesize \textsl{Workshop Scientific HPC in the pre-Exascale era (part of ITADATA2024)}}
\usepackage{fancyhdr}
\fancypagestyle{empty}{\fancyhf{}\fancyhead[C]{\itadata}
  \fancyhead[R]{\includegraphics[width=.05\textwidth]{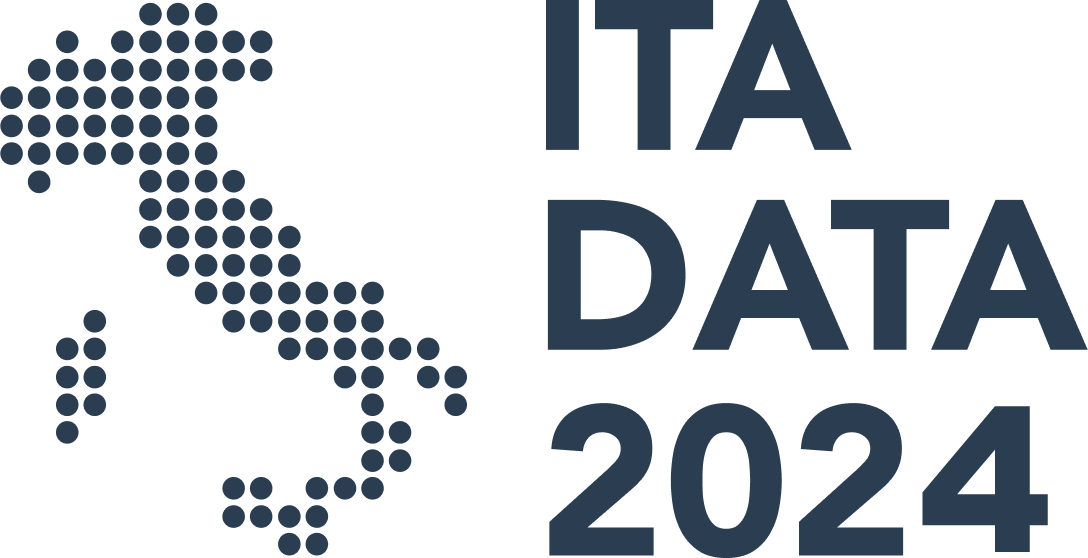}}}

\usepackage{amssymb}
\usepackage{array}

\makeatletter
\begin{document}
\title{The Gaia AVU--GSR solver: a CPU + GPU parallel code toward Exascale systems}
\author{Valentina Cesare\inst{1}\orcidID{0000-0003-1119-4237} \and
Ugo Becciani\inst{1}\orcidID{0000-0002-4389-8688} \and
Alberto Vecchiato\inst{2}\orcidID{0000-0003-1399-5556} \and
Mario Gilberto Lattanzi\inst{2}\orcidID{0000-0003-0429-7748} \and
Marco Aldinucci\inst{3}\orcidID{0000-0001-8788-0829} \and
Beatrice Bucciarelli\inst{2}\orcidID{0000-0002-5303-0268}
}
\authorrunning{V. Cesare et al.}
\institute{INAF, Astrophysical Observatory of Catania, via Santa Sofia 78, 95123 Catania, CT, Italy \email{valentina.cesare@inaf.it} \and
INAF, Astrophysical Observatory of Turin, via Osservatorio 20, 10025 Pino Torinese, TO, Italy \and
University of Turin, Computer Science Department, corso Svizzera 185, 10149 Turin, TO, Italy}
\maketitle              \begin{abstract}
The solver module of the Astrometric Verification Unit -- Global Sphere Reconstruction (AVU--GSR) pipeline aims to find the astrometric parameters of $\sim$$10^8$ stars in the Milky Way, besides the attitude and instrumental settings of the Gaia satellite and the parametrized post Newtonian parameter $\gamma$ with a resolution of 10-100 micro-arcseconds. To perform this task, the code solves a system of linear equations with the iterative Least Squares (LSQR) algorithm, where the coefficient matrix is large (10-50 TB) and sparse and the iterations stop when convergence is reached in the least squares sense. The two matrix-by-vector products performed at each LSQR step were GPU-ported, firstly with OpenACC and then with CUDA, resulting in a $\sim$$1.5$x and $\sim$$14$x speedup, respectively, over an original code version entirely parallelized on the CPU with MPI + OpenMP. The CUDA code was further optimized and then ported with programming frameworks portable across different GPU architectures, obtaining a further $\sim$$2$x acceleration factor. One critical section of the code consists in the computation of covariances, whose total number is $N_{\rm unk} \times (N_{\rm unk} - 1)/2$ and occupy $\sim$1 EB, being Nunk $\sim$$5 \times 10^8$ the total number of unknowns. This ``Big Data'' issue cannot be faced with standard approaches: we defined an I/O-based pipeline made of two concurrently launched jobs, where one job, i.e., the LSQR, writes the files and the second job reads them, iteratively computes the covariances and deletes them. The pipeline does not present significant bottlenecks until a number of covariances elements equal to $\sim$$8 \times 10^6$. The code currently runs in production on Leonardo CINECA infrastructure.

\end{abstract}
\section{Introduction}
\label{sec:Introduction}

The size of the data produced by new scientific experiments, such as the Square Kilometer Array (SKA)~\cite{Dewdney_SKA_2009} and Euclid~\cite{Euclid_Collaboration_2024}, is rapidly increasing, also occupying $\sim$$10^2$ PB in storage. To analyse these data in human-sized timescales, a specific co-design between hardware and software is needed. Heterogeneous architectures represent the mainstream~\cite{Carpenter_2022}, where one or more accelerators, such as GPUs or FPGAs, are linked with high-bandwidth bridges (e.g., NVLINK with $\sim$300 GB/s\footnote{\url{https://www.nvidia.com/en-us/design-visualization/nvlink-bridges/}}) to the host (CPU) of each node and the different nodes, about $\sim$$10^3$, are interconnected with high-bandwidth and low-latency networks (e.g., Mellanox/Cray with $\gtrsim$200 Gb/s~\cite{Pearson_2023}). To optimally exploit this hardware, the software have to be properly designed and written with specific parallelization frameworks, such as, OpenACC, OpenCL, SYCL, OpenMP with GPU offload, HIP, and C++ Parallel Standard Template Library (PSTL) (e.g.,~\cite{Malenza_2024}).

As an example of scientific application which exploits the mentioned hardware and software techniques and deals with large data volumes, we present the solver module of the Astrometric Verification Unit -- Global Sphere Reconstruction (AVU--GSR) parallel pipeline for the European Space Agency (ESA) Gaia mission\footnote{\url{https://sci.esa.int/web/gaia}}~\cite{Becciani_2014}, developed under the Data Processing and Analysis Consortium (DPAC)~\cite{Vecchiato_2018} and managed by the Data Processing Centre of Torino (DPCT). The code exploits both CPU and GPU resources, is parallelized across the nodes with MPI and within each node with CUDA, on the GPU, and has some sections with heavy I/O operations. The AVU--GSR solver aims to find the astrometric parameters of $\sim$$10^8$ primary stars in our Galaxy, besides the attitude and instrumental settings of the Gaia satellite, and the Parametrized Post-Newtonian (PPN) parameter $\gamma$~\cite{Vecchiato_2003}, with a high accuracy between 10 and 100 micro-arcseconds, under two fully relativistic astrometric models~\cite{Bertone_2017,Crosta_2017}. The maximum input size expected at the end of the Gaia mission is of $\sim$10-50 TB. The presented computational problem is an example of ``inverse problem''~\cite{Tarantola_2005}, which consists in deriving the parameters of a given model from a set of observational data and represents a use case faced in diverse scientific contexts~\cite{Borriello_1986,VanderMarel_1988,Baur_and_Austen_2005,Liu_2006,Huang_2012,Huang_2013,Naghibzadeh_and_vanderVeen_2017,Joulidehsar_2018,LIANG_2019,LSQR_geology_2019,Ling_2019,Bin_2020,Jaffri_2020,Guo_2021}. To derive the parameters, the code iteratively solves a system of linear equations with Least Squares (LSQR) algorithm~\cite{Paige_and_Saunders_1982a,Paige_and_Saunders_1982b}, to solve large, ill-posed, overdetermined, and possibly sparse systems of equations. Besides the system solution, the code calculates the variances and covariances of the system.

The code has run in production on CINECA infrastructures since 2014, under an agreement between INAF and CINECA and with the support of the Italian Space Agency (ASI). Since the second half of 2023, it has run on Leonardo CINECA infrastructure\footnote{\url{https://leonardo-supercomputer.cineca.eu/hpc-system/}}.

The largest fraction of the execution time at each LSQR step is given by three sections, i.e., two matrix-by-vector products, computed on the GPU, and a writing phase, where several files necessary for covariances calculation are saved in an output directory. Nonetheless, it is important to point out that the writing operations do not occur every LSQR iteration but every certain number of LSQR iterations, set at runtime. The problem faced by the Gaia mission presents natural correlations. This results in the computation of a large-scale system with a big-sized variance-covariances matrix, which can entail a Big Data problem. The calculation of covariances represent a nontrivial issue which might involve PB-EB of memory and requires nonconventional strategies to be solved (e.g.,~\cite{Eftekhari_2018,Dumitru}). 

The outline of this work is as follows. After a general description of the Gaia AVU--GSR solver (Section~\ref{sec:Gaia_AVU-GSR}), we describe the two most computationally important sections of the solver, i.e., the LSQR in general (Section~\ref{sec:Gaia_AVU-GSR-LSQR}) and the computation of covariances performed through a two programs pipeline (Section~\ref{sec:Gaia_AVU-GSR-Covariances_Pipeline}). Section~\ref{sec:Results} presents the tests executed to measure the performance of the covariances pipeline and Section~\ref{sec:Conclusions} concludes the paper.

\section{Structure of the Gaia AVU--GSR solver}
\label{sec:Gaia_AVU-GSR}

To find the astrometric, attitude, instrumental, and global parameters, the AVU--GSR solver iteratively solves, through a customized and preconditioned version of the LSQR algorithm, an overdetermined system of linear equations,
\begin{equation}
    \label{eq:System}
    \mathbf{A} \times \vec{x} = \vec{b},
\end{equation}
under two relativistic astrometric models~\cite{Bertone_2017,Crosta_2017}. The quantities in Eq.~\eqref{eq:System} are: (1) the large and sparse coefficient matrix $\mathbf{A}$, expected to contain, at mission end, $\sim$$10^{11} \times (5 \times 10^8)$ elements of double type ($\sim$$10^{11}$ rows and $\sim$$5 \times 10^8$ columns), corresponding to $\sim$400 EB; (2) the unknowns array $\vec{x}$ with $\sim$$5 \times 10^8$ elements of double type, corresponding to $\sim$4 GB; (3) the known terms array $\vec{b}$ with $\sim$$10^{11}$ elements of double type, corresponding to $\sim$800 GB. In matrix $\mathbf{A}$, the rows, equal in number to the elements of $\vec{b}$, represent the equations of the system and the observations of the stars, which are $\sim$$10^3$ per star. Instead, the number of columns of $\mathbf{A}$, equal to the elements of $\vec{x}$, represents the number of the system unknowns. For reasons of computational time, calculations are not performed with the sparse coefficient matrix but with a dense matrix, which only contains, at maximum, 24 nonzero elements per row, $N_{\rm Astro} = 5$, $N_{\rm Att} = 12$, $N_{\rm Instr} = 6$, and $N_{\rm Glob} = 1$, for the astrometric, attitude, instrumental, and global parts, respectively. Therefore, the dense matrix will reduce its maximum number of elements from $\sim$$10^{11} \times (5 \times 10^8)$ to $\sim$$10^{11} \times 24$ and its maximum size from $\sim$400 EB to $\sim$19 TB.

The number of equations in the system is larger than the number of unknowns, which makes the system overdetermined and makes a set of constraints equations to be defined for a univocal solution. The LSQR iterates until a least squares convergence condition or when a maximum number of iterations set at runtime is achieved.

In the LSQR, two matrix-by-vector products are computed per iteration, i.e., the {\it aprod} 1,
\begin{equation}
    \label{eq:aprod_1}
    \vec{b}^{itn} += \mathbf{A} \times \vec{x}^{itn-1},
\end{equation}
which provides an iterative estimate of the known terms, and the {\it aprod} 2,
\begin{equation}
    \label{eq:aprod_2}
    \vec{x}^{itn} += \mathbf{A}^T \times \vec{b}^{itn},
\end{equation}
which provides an iterative estimate of the solution.

The overall solver module is structured in the following way:
\begin{enumerate}
    \item Input data are imported from external files in binary format;
    \item The system is preconditioned;
    \item Initial conditions are set with {\it aprod} 2 (Eq~\eqref{eq:aprod_2});
    \item LSQR while loop (up to convergence condition or maximum number of iterations):
    \begin{enumerate}
        \item The known terms array $\vec{b}$ is iteratively computed with {\it aprod} 1 (Eq~\eqref{eq:aprod_1});
        \item The solution array $\vec{x}$ is iteratively computed with {\it aprod} 2 (Eq~\eqref{eq:aprod_2});
        \item System variances are iteratively computed;
        \item System covariances are iteratively computed (if the AVU--GSR solver is launched without the $-noCov$ option);
    \end{enumerate}
    \item The system is de-preconditioned;
    \item The system solution is saved to files in binary format.
\end{enumerate}

In the two following subsections, the structure and the parallelization of the LSQR in general and of its section dedicated to covariances calculation are detailed.

\subsection{The LSQR}
\label{sec:Gaia_AVU-GSR-LSQR}

The LSQR represents the $\sim$95\% of the computational time of the solver module and $\gtrsim$90\% of the computational time of each LSQR iteration is provided by the two matrix-by-vector products of {\it aprod} 1 and {\it aprod} 2. 

The system of equations (Eq.~\eqref{eq:System}), solved with LSQR algorithm, is parallelized over many computational nodes with MPI, where the number of required nodes is set according to the input memory of the system. Specifically, the computation related to a subset of the total number of observations (rows in matrix $\mathbf{A}$), is assigned to a different MPI process~\cite{Becciani_2014,Cesare_2021,Cesare_INAF_Technical_Report_OpenACC_163_2022,Cesare_INAF_Technical_Report_CUDA_164_2022,Cesare_OpenACC_2022,Cesare_CUDA_2023,Malenza_2024}. In the coefficient matrix $\mathbf{A}$, the computation related to the astrometric part, representing $\sim$90\% of the entire system, is distributed over the MPI processes, whereas the attitude, instrumental, and global parts are replicated on each MPI process. Also the computation referred to the constraints equations, $\sim$1\% of the entire system, is replicated among the MPI processes.

In the AVU--GSR solver, the fraction of the total execution time due to MPI communications is $\sim$10\%, i.e., it is subdominant compared to computation~\cite{Malenza_2024}. The most important MPI communications are two reduce operations per LSQR iteration, performed at the end of each {\it aprod} 1 and {\it aprod} 2 call to combine the known terms and solution results among the different MPI processes.

The sections of code assigned to each MPI process were ported to the GPU. In a first version of the code, the intra-MPI process parallelization was managed on the CPU with OpenMP~\cite{Becciani_2014}. The code was successively GPU-ported, replacing OpenMP with OpenACC~\cite{Cesare_INAF_Technical_Report_OpenACC_163_2022,Cesare_OpenACC_2022}, obtaining a $\sim$$1.5$x speedup over the OpenMP version, and, then, replacing OpenACC with CUDA~\cite{Cesare_INAF_Technical_Report_CUDA_164_2022,Cesare_CUDA_2023}, obtaining a $\sim$$14$x speedup over the OpenMP version. The CUDA version was further optimized, reaching an additional $\sim$$2$x speedup~\cite{Malenza_2024}. This version will soon officially replace the first CUDA version in production. This optimized CUDA version was also ported with other parallel frameworks portable to different GPU architectures~\cite{Malenza_2024}. The most computational intensive regions running on the GPU are the two {\it aprod} functions, where a CUDA kernel was defined for each region of the system (astrometric, attitude, instrumental, and global), one for the {\it aprod} 1 and one for the {\it aprod} 2. The MPI processes are assigned to the GPUs of each node in a round-robin fashion and, in each CUDA kernel, the grid of GPU threads is defined to match the topology of the problem to solve for performance reasons.

To best exploit the computational resources, the solver requires the parallelization over many THIN nodes (i.e., nodes with a CPU memory of 256-512 GB since a larger CPU memory is not necessary), where the resources of the single node are maximally occupied. Leonardo, the computer cluster where the code runs in production, has 32 cores @ 2.6 GHz and 4 NVIDIA Ampère A100 of 64 GB each per node, with a total of 512 GB of CPU memory and 256 GB of GPU memory per node. Since the code runs on GPU, the maximum memory assigned per node is limited by the GPU memory of the node. For example, a system occupying a memory of 450 GB will require two instead of one node, even if it fits the CPU memory of one node.

\subsection{The covariances pipeline}
\label{sec:Gaia_AVU-GSR-Covariances_Pipeline}

Whereas the system variances, that quantify the errors on the solutions, are calculated in each production run, the system covariances~\cite{Cesare_SPIE_Covariances_2024}, that quantify the correlations between two unknowns, have not been calculated yet. The calculation of covariances is defined in the solver in a section of code that is only executed if a particular flag, $-noCov$ (numbered list in Section~\ref{sec:Gaia_AVU-GSR}), is not set at runtime. We plan to calculate covariances in the next production runs and we discuss here the computational problem arisen by this calculation.

The total number of elements of the variances-covariances matrix is $N_{\rm cov} = N_{\rm unk} \times (N_{\rm unk} - 1)/2$, which is equal to $\sim$$1.25 \times 10^{17}$ since we expect $N_{\rm unk} \sim 5 \times 10^8$ at the end of the Gaia mission. Being the covariances of double precision type, we expect a memory of $\sim$1 EB occupied by the variances-covariances matrix, which cannot be computationally managed on current infrastructures. In production runs, we do not plan to compute the entire variances-covariances matrix but a subset of its elements, according to specific scientific needs. Yet, also computing this subset might not be trivial.

The covariances, $\vec{Cov}$, are computed at each LSQR iteration $itn$ with this formula:
\begin{equation}
    \label{eq:Covariances}
    Cov^{itn}[j] += factor^{itn} \cdot x^{itn}[j_1] \cdot x^{itn}[j_2],
\end{equation}
with $factor^{itn}$ an iteration-dependent multiplication factor and $(j_1,j_2)$ the couples of indexes of the elements of the solution array $\vec{x}$ between which the covariances have to be computed. Since for each covariance we require two indexes, the total number of indexes for the covariances to compute is $N_{\rm cov, indexes} = 2 \times N_{\rm cov}$. It is important to point out that $j_1$ and $j_2$ are global indexes but $\vec{x}$ is a local array, defined on each MPI process. If to evaluate expression~\eqref{eq:Covariances} each MPI process would have to broadcast $\vec{x}$ to all the other MPI processes, the code would become soon MPI communications-bound and performance would parabolically drop with the number of nodes.

Therefore, we adopted a different strategy with a two-programs pipeline based on disk I/O. Concurrently to the AVU--GSR solver, a second program, which we will call ``the covariances program'' from this moment on, is launched. At the end of each LSQR iteration, after the execution of {\it aprod} 1 and {\it aprod} 2 kernels on GPU, on CPU the local content of $\vec{x}$ related to the first $itnCovCP$ iterations, where $itnCovCP$ is set at runtime, is stored by each MPI process in a 1D and double-precision local array, $\vec{x_{\rm Cycle}}$. When $itn = itnCovCP$, $\vec{x_{\rm Cycle}}$ is written to file by each MPI process in an output directory. The covariances program interrogates the filesystem at regular intervals to check if these files are present in the output directory. Once they are present, the covariances program reads the files, properly rearranging the information that contain, and calculate the corresponding covariances. Once these operations are concluded, the covariances program deletes the files. Meanwhile, the LSQR in the AVU--GSR solver has continued to iterate. If when $itn = 2 \times itnCovCP$ the files of the previous cycle of $itnCovCP$ iterations have not been deleted yet by the covariances program, the LSQR waits, interrogating the filesystems at regular intervals until the files of the previous cycle are not deleted. Therefore, for an efficient pipeline, the reading + covariances calculation phase of $itnCovCP$ iterations in the covariances program has to be shorter than the LSQR + writing phase of $itnCovCP$ iterations in the AVU--GSR solver, to avoid waiting times bottlenecks.

The AVU--GSR solver and the covariances program run on different computational resources. Whereas the AVU--GSR solver runs on multiple nodes, the covariances program sequentially runs on a single node. Therefore, if the solver code needs to run on $N$ nodes for a given input dataset, the total number of nodes required for a run is $N + 1$. Moreover, the AVU--GSR solver executes on GPU resources, excluding the regions dedicated to covariances, whereas the covariances program entirely runs on CPU.

We implemented this pipeline in two different versions. In version n. 1, each MPI process writes one file every $itnCovCP$ iterations with the local information of $\vec{x}$ referred to these iterations. The content of $\vec{x}$ referred to every cycle of $itnCovCP$ iterations is saved in the $\vec{x_{\rm Cycle,1}}$ array, which is written to file. Being the size of $\vec{x}$ equal to $\sim$$N_{\rm unk}/N_{\rm proc}$, where $N_{\rm proc}$ is the number of MPI processes, the total size of the files written per cycle of $itnCovCP$ iterations is:
\begin{equation}
    \label{eq:Size_Cycle_1}
    Size_{{\rm Cycle}, 1} = 8 \times itnCovCP \times N_{\rm unk},
\end{equation}
where the factor of 8 is due to the fact that $\vec{x_{\rm Cycle,1}}$ array is of double type.

In version n. 2, each MPI process writes two files every $itnCovCP$ iterations with only the local information of $\vec{x}$, referred to these iterations, related to the covariances to compute, i.e., correspondent to the global indexes $j_1$ and $j_2$ of Eq.~\eqref{eq:Covariances}. The content of the two files is stored in two arrays, $\vec{x_{\rm Cycle,2.1}}$ and $\vec{x_{\rm Cycle,2.2}}$. Each of these two arrays contain a number of elements equal to $\sim$$itnCovCP \times N_{\rm cov. indexes}/(2 \times N_{\rm proc})$ and, thus, the total size of the files written per cycle of $itnCovCP$ iterations is:
\begin{equation}
    \label{eq:Size_Cycle_2}
    Size_{{\rm Cycle}, 2} = 8 \times itnCovCP \times N_{\rm cov. indexes}.
\end{equation}

Looking at Eqs.~\eqref{eq:Size_Cycle_1} and~\eqref{eq:Size_Cycle_2}, we can see that, in version n. 1, the writing phase in the LSQR only depends on the total number of unknowns in the system, $N_{\rm unk}$, and not on the number of covariances indexes to calculate, $N_{\rm cov. indexes}$. The opposite situation occurs in version n. 2. This indicates that, until $N_{\rm cov. indexes} < N_{\rm unk}$, the writing phase in version n. 2 can be more efficient than in version n. 1 and the situation can reverse when $N_{\rm cov. indexes} > N_{\rm unk}$. In version n. 1, how many and which covariances to calculate is set in the covariances program, whereas, in version n. 2, it is already set in the AVU--GSR solver.

Table~\ref{tab:Differences_Versions_Covariances_Pipeline} summarizes the differences between the two versions of the pipeline.

\begin{table}
\caption{Summary of the differences between the two versions of the AVU--GSR solver + covariances program pipeline.}\label{tab:Differences_Versions_Covariances_Pipeline}
\begin{tabular}{|p{5.9cm}|p{5.9cm}|}
\hline
\textbf{Version n. 1} & \textbf{Version n. 2} \\
\hline
Each MPI process writes one file every $itnCovCP$ iterations with the local information of $\vec{x}$ referred to these iterations. & Each MPI process writes two files every $itnCovCP$ iterations with only the local information of $\vec{x}$, referred to these iterations, related to the covariances to compute.\\
\hline
How many and which covariances to calculate is set in the covariances program. & How many and which covariances to calculate is already set in the AVU--GSR solver.\\
\hline
The total size of the files written every $itnCovCP$ iterations is given by Eq.~\eqref{eq:Size_Cycle_1} and only depends on the number of unknowns of the system ($N_{\rm unk}$). & The total size of the files written every $itnCovCP$ iterations is given by Eq.~\eqref{eq:Size_Cycle_2} and only depends on the number of covariances indexes ($N_{\rm cov. indexes}$). \\
\hline
If $N_{\rm cov. indexes} > N_{\rm unk}$, the writing phase in version n. 1 can be more efficient than in version n. 2. & If $N_{\rm cov. indexes} < N_{\rm unk}$, the writing phase in version n. 2 can be more efficient than in version n. 1. \\

\hline
\end{tabular}
\end{table}

\section{Results}
\label{sec:Results}

We tested the performance of the AVU--GSR solver + covariances program pipeline (which we will call ``covariances pipeline'' for simplicity from this moment on) by evaluating the execution time of a cycle of $itnCovCP$ iterations in the AVU--GSR solver and in the covariances program. In the AVU--GSR solver, a cycle of $itnCovCP$ iterations includes $itnCovCP - 1$ ``standard'' iterations, where {\it aprod} 1 and {\it aprod} 2 are executed, variances are calculated, and the $\vec{x_{\rm Cycle}}$ array is filled iteration per iteration, and an iteration where besides these operations the files needed for covariances calculation are written. We will call it the ``Iter + Write'' phase and its execution time $t_{\rm Iter + Write,Cycle}$. In the covariances program, a cycle of $itnCovCP$ iterations includes the reading of the written files of the current cycle and the calculation of the corresponding covariances. We will call it the ``Read + Cov'' phase and its execution time $t_{\rm Read + Cov,Cycle}$. In our tests, $t_{\rm Iter + Write,Cycle}$ and $t_{\rm Read + Cov,Cycle}$ are averaged among the different cycles of $itnCovCP$ iterations. The execution times $t_{\rm Iter + Write,Cycle}$ and $t_{\rm Read + Cov,Cycle}$ determine the speed of the pipeline because until $t_{\rm Read + Cov,Cycle} \leq t_{\rm Iter + Write,Cycle}$ no waiting times occur in the LSQR and the pipeline can proceed; if $t_{\rm Read + Cov,Cycle} > t_{\rm Iter + Write,Cycle}$ a substantial bottleneck can occur. We compare the performance of versions n. 1 and n. 2 of the pipeline, to see which one is more efficient.

We run these performance tests on Leonardo CINECA infrastructure since it is the platform that we use for the production of the AVU--GSR solver. Leonardo has two storage areas, a capacity tier, that we adopt for our performance tests, and a fast tier. The capacity tier has a total capacity of 106 PB, made of 31 $\times$ DDN EXAScaler SFA7990X appliances for HDD storage and 4 $\times$ DDN EXAScaler SFA400NVX for metadata, and a bandwidth of 620 GB/s. The fast tier has a total capacity of 5.4 PB, made of 31 $\times$ DDN Exascaler ES400NVX2, is completely full flash and based on NVMe and SSD disks and has a bandwidth of 1.4 TB/s.

To run the performance tests, we used synthetic and randomly generated input data, distributed in the system as the real data. This choice was led by the fact that the size of the synthetic input data ($Mem$) can be passed as a runtime parameter in GB, and, thus, it can be directly controlled by the programmer according to specific needs. The generated system has $N_{\rm Astro} = 5$, $N_{\rm Att} = 12$, $N_{\rm Instr} = 6$, and $N_{\rm Glob} = 0$ 
per row, where the global part has not been considered so far in production runs. If run up to convergence, the solver would find the identity solution with the synthetic data as input; yet, we set a LSQR iteration limit of 100, which is sufficient to get a significant statistics.

To study the performance of the two versions of the pipeline, the parameter space can be explored across three different axes:
\begin{enumerate}
    \item $N_{\rm unk}$: varying $N_{\rm unk}$ allows to explore how the duration of the Iter + Write phase changes in version n. 1 of the pipeline.
    \item $N_{\rm cov. indexes}$: varying $N_{\rm cov. indexes}$ allows to explore how the duration of the Iter + Write phase changes in version n. 2 of the pipeline.
    \item $itnCovCP$: both versions of the pipeline depend on this parameter and varying it allows, besides to explore how it affects the performance of the overall pipeline, to investigate the optimal trade-off between too frequent accesses to the filesystem and the writing of too large files. In this work, we do not investigate how the pipeline is influenced by the variation of $itnCovCP$, which is set to 20. Running the solver for 100 iterations, we have 5 writing + reading cycles. The variation of $itnCovCP$ will be object of a future work.
\end{enumerate}

The way in which the solver is built does not allow to directly act on $N_{\rm unk}$ parameter. Indeed, setting $Mem$ at runtime, a system with $N_{\rm obs}$ and $N_{\rm unk}$ set such that the system occupies $Mem$ GB is generated but a system with a double input $Mem$ does not necessarily have a double $N_{\rm unk}$~\cite{Cesare_SPIE_Covariances_2024}. Therefore, at this stage of our analysis, we investigate a scalability along the $Mem$, instead of $N_{\rm unk}$, axis. 

\begin{figure}
    \includegraphics[width=\textwidth]{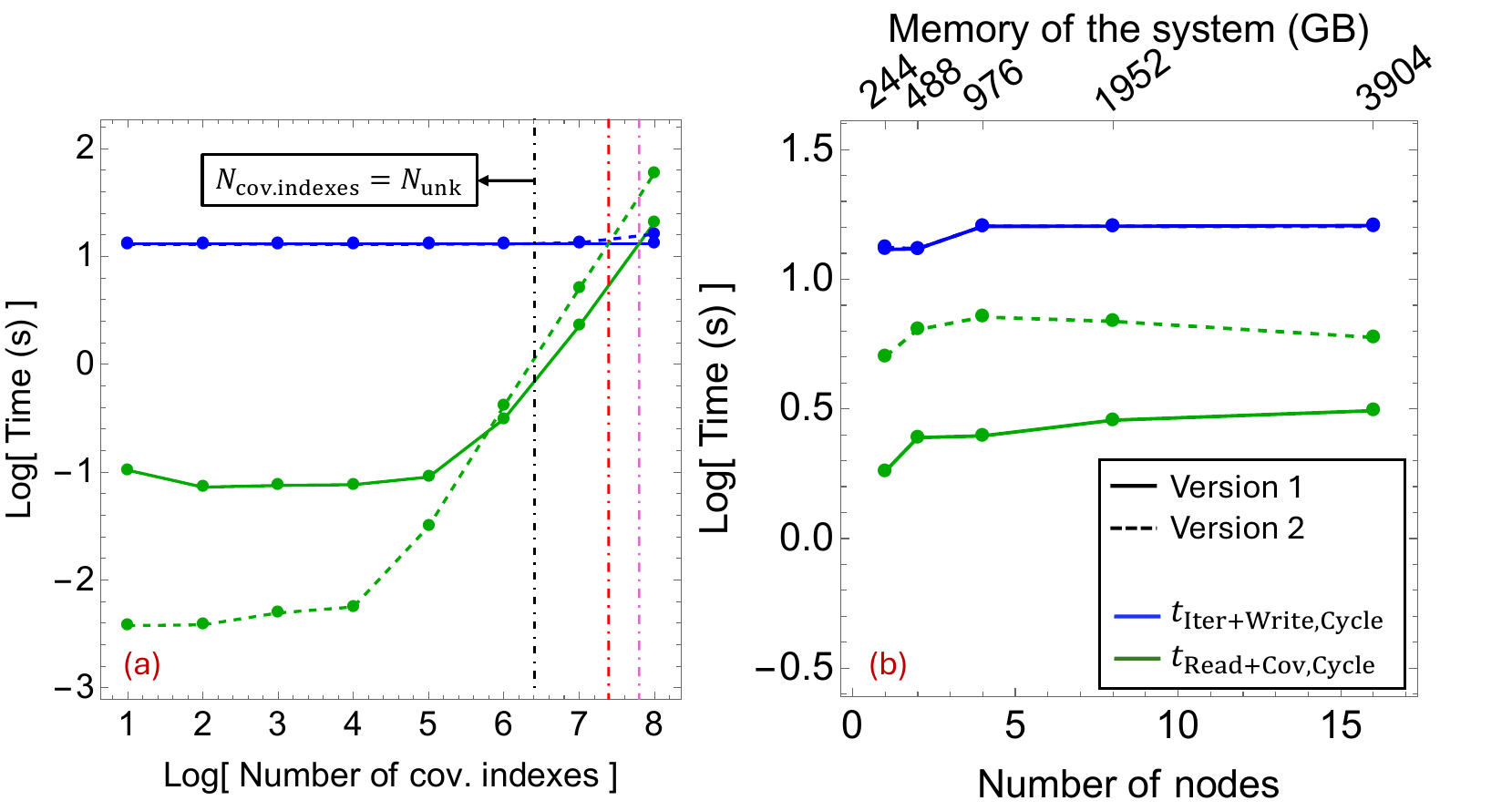}
    \caption{Performance tests for the AVU--GSR solver + covariances program pipeline. Figures~\ref{fig:Performance_Tests_Covariance}a and~\ref{fig:Performance_Tests_Covariance}b show tests n. 1 and n. 2, respectively. The solid and dashed lines refer to versions n. 1 and n. 2 of the pipeline in both panels. The blue and green lines represent the time of the Iter + Write phase ($t_{\rm Iter + Write, Cycle}$) and the Read + Cov phase ($t_{\rm Read + Cov, Cycle}$) in both panels. In Figure~\ref{fig:Performance_Tests_Covariance}a, the vertical dot-dashed black, red, and pink lines refer to the $N_{\rm cov. indexes} = N_{\rm unk} \sim 2.55 \times 10^6$, $N_{\rm cov. indexes} = 1.6 \times 10^7$ and $N_{\rm cov. indexes} = 6.3 \times 10^7$ points, respectively. This figure is readapted from Figure 1 in~\cite{Cesare_SPIE_Covariances_2024}.} \label{fig:Performance_Tests_Covariance}
\end{figure}

To investigate the performance of the pipeline, we performed two different tests. In test n. 1 (Figure~\ref{fig:Performance_Tests_Covariance}a), we run the AVU--GSR solver on 4 MPI processes of one node of Leonardo, we set $Mem = 244$ GB, to nearly maximize the GPU occupancy of the node (95\%), and we varied $N_{\rm cov. indexes}$ from $10^1$ to $10^8$ at intervals of 1 dex, which means that $N_{\rm cov}$ is varied from $5 \times 10^0$ to $5 \times 10^7$ at intervals of 1 dex. The system of $Mem = 244$ GB has $N_{\rm unk} = 2551805 \sim 2.55 \times 10^6$. In test n. 2 (Figure~\ref{fig:Performance_Tests_Covariance}b), we set $N_{\rm cov. indexes} = 10^7$ ($N_{\rm cov} = 5 \times 10^6$), we run the AVU--GSR solver on a number of nodes, $N_{\rm nodes}$, of Leonardo from 1 to 16, with 4 MPI processes per node, and we set $Mem = (244 \text{ GB}) \times N_{\rm nodes} = \{244, 488, 976, 1952, 3904\} \text{ GB}$ on $N_{\rm nodes} = \{1,2,4,8,16\}$. The maximum size occupied by the input dataset, $Mem = 3904$ GB, is $\sim$40\% of the size that we expect at the end of the Gaia mission.

In both panels of Figure~\ref{fig:Performance_Tests_Covariance}, the solid and dashed lines refer to versions n. 1 and n. 2 of the pipeline, respectively. In Figure~\ref{fig:Performance_Tests_Covariance}a, the black dot-dashed vertical line is the $N_{\rm cov. indexes} = N_{\rm unk}$ line as a reference. The blue lines represent the Iter + Write time in the AVU--GSR solver and the green lines the Read + Cov time in the covariances program. 

In Figure~\ref{fig:Performance_Tests_Covariance}a, we can see that the blue line of version n. 1 (blue solid line) presents a perfect horizontal trend since the Iter + Write time does not depend at all on $N_{\rm cov. indexes}$. However, also the Iter + Write time in version n. 2 presents a weak dependence on $N_{\rm cov. indexes}$: the ratio between $t_{\rm Iter + Write, Cycle}$ times in versions n. 2 and n. 1 is $\sim$$1$ up to $N_{\rm cov. indexes} \sim 10^7$ and when $N_{\rm cov. indexes} = 10^8$ it only rises to $1.23$. This is due to the fact that the dependence on $N_{\rm cov. indexes}$ in version n. 2 is only present in the writing phase, which occurs every $itnCovCP$ iterations. This result is obtained for $itnCovCP = 20$ and future studies aim to investigate if it is preserved for smaller and larger values of $itnCovCP$. It is expected that the smaller $itnCovCP$, the stronger the dependence of $t_{\rm Iter + Write, Cycle}$ time in version n. 2 on $N_{\rm cov. indexes}$, because the writing phases, which depend on $N_{\rm cov. indexes}$, occur more frequently.

The $t_{\rm Read + Cov, Cycle}$ time in version n. 1 is larger than $t_{\rm Read + Cov, Cycle}$ in version n. 2 up to $N_{\rm cov. indexes} \sim 10^6$, slightly before the $N_{\rm cov. indexes} = N_{\rm unk}$ line. The situation reverses when $N_{\rm cov. indexes} \gtrsim 10^6$. It is important to point out that until $N_{\rm cov. indexes}$ $\lesssim$ $1.6 \times 10^7$ ($N_{\rm cov} \lesssim 8 \times 10^6$) $t_{\rm Read + Cov, Cycle} < t_{\rm Iter + Write, Cycle}$ for both pipeline versions. This point is highlighted in Figure~\ref{fig:Performance_Tests_Covariance}a with a red dot-dashed vertical line. Up to this point the two versions of the pipeline are efficient, since there are not bottlenecks due to waiting times for the deletion of the files of the previous cycle of $itnCovCP$ iterations, and are basically equally performant, since the speed of the pipeline is entirely driven by $t_{\rm Iter + Write, Cycle}$ which is basically coincident between the two pipeline versions. For $1.6 \times 10^7 \lesssim N_{\rm cov. indexes} \lesssim 6.3 \times 10^7$ ($8.0 \times 10^6 \lesssim N_{\rm cov} \lesssim 3.15 \times 10^7$), version n. 1 of the pipeline continues to be efficient because its $t_{\rm Read + Cov, Cycle}$  is smaller than $t_{\rm Iter + Write, Cycle}$ but the same does not hold anymore for version n. 2 of the pipeline. The $N_{\rm cov. indexes} = 6.3 \times 10^7$ point is highlighted with a vertical pink dot-dashed line. For $N_{\rm cov. indexes} \gtrsim 6.3 \times 10^7$ ($N_{\rm cov} \gtrsim 3.15 \times 10^7$), both pipeline versions are not efficient anymore.

Figure~\ref{fig:Performance_Tests_Covariance}b shows that version n. 1 and n. 2 of the pipeline are equally efficient along the entire range of $N_{\rm nodes}$ (bottom axis) and $Mem$ (top axis). Indeed, although $t_{\rm Read + Cov, Cycle}$ in version n. 2 is larger than the corresponding time in version n. 1 along the entire $N_{\rm nodes}$ and $Mem$ ranges, $t_{\rm Iter + Write, Cycle}$ in the two pipelines versions basically coincide and, therefore, the pipelines' speeds are about the same. As said before, varying $Mem$ proportional to $N_{\rm nodes}$ does not necessarily imply than the same proportionality is observed by $N_{\rm unk}$. Specifically, for $Mem = \{244, 488, 976, 1952, 3904\}$ GB, we have $N_{\rm unk} = \{2.55, 4.15,$ $4.15, 7.35, 13.8\} \times 10^6$. Yet, $t_{\rm Iter + Write, Cycle}$ in version n. 1 presents anyway a quite good weak scaling trend. This is due to the fact that only the writing phase depends on $N_{\rm unk}$ alone but the other operations in the LSQR iterations, mainly {\it aprod} 1 and {\it aprod} 2, depend on the entire $Mem$, which also explain the fact that the same trend is observed in $t_{\rm Iter + Write, Cycle}$ in version n. 2. A different situation occurs in test n. 1, where the Iter + Write phase in version n. 1 does not depend at all on $N_{\rm cov. indexes}$ and, therefore, its trend as a function of $N_{\rm cov. indexes}$ is perfectly horizontal. 

Summarizing, since in test n. 1 $t_{\rm Read + Cov, Cycle}$ $\lesssim$ $t_{\rm Iter + Write, Cycle}$ along nearly the entire range of $N_{\rm cov. indexes}$ and in test n. 2 $t_{\rm Read + Cov, Cycle}$ $\lesssim$ $t_{\rm Iter + Write, Cycle}$ along the entire range of $N_{\rm nodes}$ and $Mem$, we can conclude that the two pipeline versions are about equally efficient. Therefore, which one would we prefer? In future runs, we aim to compute a substantial number of covariances which could often be larger than $N_{\rm unk}$, which leads our choice to version n. 1. Moreover, there is another advantage that version n. 1 provides over version n. 2. We do not plan to compute covariances for every production run but only for a selected number of runs. Therefore, we might decide not to concurrently run the AVU--GSR solver and the covariances program but to firstly run the AVU--GSR solver, maintaining all the files, and, in a next stage, to run the covariances program without deleting the files cycle by cycle. With version n. 2 of the pipeline, we would have to decide prior to the execution of the AVU--GSR solver which and how many covariances to calculate, which makes the output files tailored for a specific case. This is no more true for version n. 1 of the pipeline, where the AVU--GSR solver can be executed and only at a second stage, after careful scientific considerations, we can decide which are the covariances that are necessary to compute. Moreover, with version n. 1 of the pipeline, we could even evaluate to compute different sets of covariances in successive runs from the same output files of the AVU--GSR solver. 

Preserving all the files in the output directory would not necessarily imply a storage issue both in version n. 1 and in version n. 2 of the covariances pipeline. Given Eqs.~\eqref{eq:Size_Cycle_1} and~\eqref{eq:Size_Cycle_2}, the size of all the files written per cycle of $itnCovCP = 20$ in the run on 16 nodes, which occupies about $40$\% of the total input size expected at the end of the Gaia mission, is of $Size_{\rm Cycle,1} = 2.2$ GB for version n. 1 and of $Size_{\rm Cycle,1} = 1.6$ GB for version n. 2 with $N_{\rm cov. indexes} = 10^7$. Considering a number of iterations typical for convergence of $N_{\rm iter} \sim 3 \times 10^5$, the total size occupied by all the files at the end of the run if they are not deleted cycle by cycle is of 33 TB in version n. 1 and of 24 TB in version n. 2, which only represent $\sim$$10^{-4}$ of the storage capacity of the Leonardo's capacity tier (106 PB) and $\sim$$5 \times 10^{-3}$ of the total storage capacity of the Leonardo's fast tier ($5.4$ PB). Only when all the needed covariances for a specific run will be computed, we will delete the output files.

\section{Conclusions}
\label{sec:Conclusions}

We described the solver module of the Gaia AVU--GSR pipeline to find the astrometric parameters of $\sim$$10^8$ primary stars in our Galaxy with a high accuracy at the 10-100 micro-arcseconds level. To find these parameters, the solver executes a customized version of the iterative LSQR algorithm to solve a system of linear equations with a maximum input size of 10--50 GB expected at the end of the Gaia mission. The LSQR part was ported to GPU with CUDA, obtaining a speedup of $\sim$$14$x over a previous CPU version of the code, which rised to $\sim$$28$x after a further optimization of the CUDA code. In May 2022, the first CUDA version of the code entered in production on Marconi100 and, at the end of 2023, the code migrated its production from Marconi100 to Leonardo CINECA infrastructure.

A section of the code at the end of each LSQR iteration, not executed in production runs so far, is dedicated to the computation of the covariances of the system. With a number of the system unknowns $N_{\rm unk} \sim 5 \times 10^8$, we would have $N_{\rm cov} \sim 1.25 \times 10^{17}$ covariances elements, which occupy $\sim$1 EB of memory. However, we do not aim to compute the entire variances-covariances matirx in production runs but only a subset of them according to specific scientific needs.

Even computing a subset of covariances requires a specific computational strategy. To compute covariances, we defined a two programs I/O-based pipeline, where a sequential program, the covariances program, is  launched concurrently to the AVU--GSR solver. In the AVU--GSR solver, at the end of each LSQR iteration, each MPI process writes, every cycle of $itnCovCP$ iterations, what is needed to compute covariances, i.e., the local array $\vec{x}$. The covariances program starts when the files related to the first cycle of $itnCovCP$ iterations are written and, every cycle of $itnCovCP$ iterations, reads the files, calculates the corresponding covariances, and deletes the files. The LSQR continues its execution when the files of the previous cycle are deleted by the covariances program. 

We built two versions of the pipeline, where each one writes a different piece of information every cycle of $itnCovCP$ iterations. In version n. 1, the total size of the files written per cycle only depends on the number of unknowns of the system, $N_{\rm unk}$ and in version n. 2, the total size of the files written per cycle only depends on the number of covariances indexes, $N_{\rm cov. indexes}$. If the files do not occupy an excessive size, we might evaluate to firstly run the AVU--GSR solver and, then, the covariances program, without deleting the files at every cycle. This is particularly useful using version n. 1 of the pipeline, where the output files are independent on the covariances to calculate and different sets of covariances can be computed from the same output files. For a system of $\sim$$4$TB, occupying $\sim$40\% of the maximum size expected at the end of the Gaia mission, which runs up to convergence, preserving all the files does not imply a storage issue, given that their total occupancy is $\sim$$10^{-4}$ of the total storage capacity of the Leonardo's capacity tier ($106$ PB) and $\sim$$5 \times 10^{-3}$ of the total storage capacity of the Leonardo's fast tier ($5.4$ PB).

We tested the performance of the two versions of the pipeline both on one node (test n. 1) and on multiple nodes (test n. 2) of Leonardo. In test n. 1, we fixed the memory occupied by the system ($Mem = 244$ GB), which results in $N_{\rm unk} = 2.55 \times 10^6$, and varied the number of covariances indexes from $10^1$ to $10^8$ at 1 dex intervals. In test n. 2, we run from 1 to 16 nodes of Leonardo, setting $Mem = (244 \text{ GB}) \times N_{\rm Nodes}$, and fixing $N_{\rm cov. indexes} = 10^7$. In both tests, $itnCovCP = 20$. In the performance tests, we measured the times of a cycle of $itnCovCP$ iterations on the LSQR and covariances program sides ($t_{\rm Iter + Write, Cycle}$ and $t_{\rm Read + Cov, Cycle}$). For a pipeline to be efficient, $t_{\rm Read + Cov, Cycle}$ has to be smaller than $t_{\rm Iter + Write, Cycle}$, to avoid waiting times due to the files deletion. In test n. 1, this situation occurs almost along the entire range of $N_{\rm cov. idexes}$, i.e, up to $N_{\rm cov. indexes} = 1.6 \times 10^7$ ($N_{\rm cov} = 8.0 \times 10^6$) for both pipelines versions and up to $N_{\rm cov. idexes} = 6.3 \times 10^7$ ($N_{\rm cov} = 3.15 \times 10^7$) for version n. 1 alone. In test n. 2, this situation occurs along the entire range of $N_{\rm nodes}$ and $Mem$.

Further tests are planned to better investigate the performance of the AVU--GSR solver + covariances program pipeline and will be object of a future work. We aim to extend test n. 2 up to 256 nodes of Leonardo, to reach a maximum input size of $\sim$62 TB, comparable to the maximum size expected at the end of the Gaia mission. It would be interesting to verify if the nearly constant trend of $t_{\rm Iter + Write, Cycle}$ and $t_{\rm Read + Cov, Cycle}$ is maintained up to a larger number of nodes in both pipeline versions. The scaling along the $N_{\rm unk}$ instead of $Mem$ axis should be explored: for this purpose, input datasets with $N_{\rm unk}$ proportional to the number of nodes should be prepared. We also plan to explore how the performance of the pipeline changes if, fixing $N_{\rm unk}$ and $N_{\rm cov. indexes}$, $itnCovCP$ is varied. At last, the same tests should be repeated with real instead of synthetic data, to see if qualitatively similar results are obtained.

\begin{credits}
\subsubsection{\ackname} This work has been supported by the Spoke 1 ``FutureHPC \& Big-Data'' of the ICSC-Centro Nazionale di Ricerca in High Performance Computing, Big
Data and Quantum Computing and hosting entity, funded by European Union-NextGenerationEU. This work was also supported by the Italian Space Agency (ASI) (Grant No. 2018-24-HH.0) as part the Gaia mission, and by EuroHPC JU under the project EUPEX (Grant No. 101033975). The results reported in this work represent some of the targets of the two-years Mini Grant project awarded by INAF ``Investigation of Scalability, Numerical Stability, and Green Computing of LSQR-based applications involving Big Data in perspective of Exascale systems: the ESA Gaia mission case study'', of which the author VC is the PI.

\subsubsection{\discintname}
The authors have no competing interests to declare that are
relevant to the content of this article.
\end{credits}

\bibliography{report.bib} \bibliographystyle{splncs04} 

\end{document}